\documentstyle[aps,prb,multicol,psfig]{revtex}
\draft

\begin{document}
\widetext

\title{Soliton dynamics in 1D quantum antiferromagnets}
\author{A. Villares Ferrer and A. O. Caldeira} 
\address{ Instituto de F\'{\i}sica ``Gleb Wataghin''\\Departamento de F\'{\i}sica do 
Estado S\'{o}lido e Ci\^{e}ncia dos Materiais,\\Universidade Estadual de Campinas, 
13083-970, Campinas, SP, Brazil.} 
\author{A.S.T. Pires}
\address{Departamento de F\'{\i}sica, Instituto de Ci\^encias Exatas, \\
Universidade Federal de Minas Gerais, 30161-970, Belo Horizonte, MG, Brazil.}
\maketitle

\widetext
\begin{abstract}

The problem of dissipative motion of solitons in the case of tetramethyl 
ammonium manganese chloride (TMMC) is studied as a function of the external magnetic field 
and the temperature. Two specific situations are analyzed separately; the first, above the 
transition temperature $T_N$, in which the classical motion of the spin degree of freedom is 
described by a sine-Gordon (SG) equation of motion and, the second, below $T_N$, in which the 
system is described by a double sine-Gordon (2SG) equation of motion.
The existence of a dissipative regime for the soliton motion and its influence on 
the dynamical structure factor -which might be experimentally detected- are reported. 
\end{abstract}


\begin{multicols}{2}

\narrowtext
\section{Introduction}
In the past few decades it has become well-established that the physical properties 
of some magnetic materials, TMMC, CsNiF$_3$ (caesium nickel fluride) and CuCl$_2$ 2NC$_5$H$_5$
(dicloro-bis-piridine copper II), for instance, have essentially one dimensional character
above their transition temperature.\cite{MikesResu1} In those kind of materials the distance 
between magnetic ions along a given direction (magnetic chain direction) is shorter
than in the other directions. In such an arrangement the intrachain coupling constant
is typically more than two orders of magnitude stronger than the interchain 
coupling constant. Therefore, the system can be considered as a set of weakly interacting 
magnetic chains . Due to the relative
simplicity of obtaining solitonic or solitary-wave solutions in 1D systems\cite{rajara},
these quasi-one-dimensional magnets turn out to be the paradigm for the study of the influence of 
the non-liner modes (solitons) in the dynamical properties of such systems at finite 
temperatures.\cite{Holy1,Pires} Although all real magnetic materials investigated are not 
perfectly one dimensional, the assumption of the 1D behavior is shown to be in 
good agreement with the experimental results (see Ref.1 and the references therein).    
  
In magnetic materials solitons or solitary-waves can be regarded as `kinks' or `twists' in 
the spin space moving with constant speed and carrying a constant topological charge 
defined by the values of the spin variables at infinity. For low enough temperatures, when 
the linear modes (spin-waves) are not excited, the magnetic system can be represented in 
first approximation by a gas of non-interacting solitons.\cite{Schriff} Using this idea, 
Mikeska\cite{mikes1,mikes2} calculated the soliton contribution to the dynamical 
structure factor of the classical one-dimensional magnets. From both works we learn that the 
assumption of ballistic motion for solitons is the origin of the `central peak' behavior 
observed in neutron scattering experiments. A different situation could be found from the 
quantum field theory point of view when the temperature is raised. 
In this case the spin-wave (SW) modes are excited, therefore not all of the degrees of 
freedom of the system contribute to the soliton formation and a residual 
interaction (which couples the center of mass of the soliton to the spin-wave modes)
shows up. In practice, the specific form of this kind of interaction is obtained via the 
collective coordinate method\cite{christlee} in the quantization process\cite{Castro} of the
classical hamiltonian. 

The soliton-SW coupling may result in a dissipative regime to the soliton motion 
depending on the form of the potential generated by the presence of the non-linear 
excitation.\cite{desposito} As it is known, the equation of motion for the spin variable in the 
TMMC below and above $T_N$ are a 2SG and a SG equation\cite{Holy1} respectively. 
This fact makes the TMMC a suitable probe to investigate the appearance of a dissipative regime 
in the soliton motion provided that, below and above a certain N\'eel transition temperature $T_N$, 
the classical equation of motion for the spin variables are substantially different. 
The main purpose of this work will be therefore the analysis of the magnetic soliton motion
above and below $T_N$ and the possible influence of the dissipative regime, found for
$T < T_N$, on the solitonic contribution to the dynamical structure factor. In doing that, we 
will use the method developed in Ref. 9 for the analysis of the dissipative dynamics of solitons.
As pointed out before, this formalism is based on the collective coordinate method\cite{christlee} 
and allows us to transform the original hamiltonian of the spin degree of freedom into one of a 
particle (the soliton) coupled to an infinite set of linear-modes (SW).

Starting from the interacting soliton-SW hamiltonian it is possible to obtain a Brownian-like 
equation of motion for the soliton center of mass via the Feynman-Vernon formalism. This effective 
equation of motion is written in terms of a damping constant that depends on the phase shifts 
of the scattering problem that emerges by the presence of the non-linear excitation coupled to the SW. 
Therefore, the analysis of the scattering properties of the 2SG 
potential (for $T<T_N$) and the SG potential (for $T>T_N$) allows us to calculate the 
mobility as a function of the temperature and the external magnetic field. As it will be shown,
above $T_N$ the SG solitons have infinite mobility in agreement with the ballistic motion 
used to understand the neutron scattering experiment for $H/T \ge 10kOe/K$\cite{neut}. 
On the other hand, for low temperatures or low magnetic field, when the spin equation of motion 
for TMMC have a 2SG form, the soliton mobility is finite, changing the form of the 
dynamical structure factor considerably.

To begin with, in Sec. II we review the models currently applied to the spin dynamics
of the TMMC compound, above and below its transition temperature, and also the corresponding 
classical equations of motion. In Sec. III we summarize the obtainment of the quantized soliton-SW 
hamiltonian and the damping parameter of the soliton Brownian motion
is calculated as a function of the temperature and the external magnetic field.
Sec. IV is devoted to the study of the influence of the soliton damped motion on the dynamical
structure factor and, finally, our conclusions are presented in Sec. V.
  
\section{The model for TMMC}
The antiferromagnet TMMC has extensively been studied from the theoretical and experimental 
points of view. The hamiltonian describing the interacting 3D array of classical spins in
this material can be written as 
\begin{equation}
{\mathcal H}=\sum_j H_j - \frac{1}{2}J_{\bot}\sum_{i \neq i^{\prime}} \sum_{j} 
S_{i,j}S_{i^{\prime},j},
\label{Htot}
\end{equation}
where
\begin{equation}
H_j=\sum_{k}\left\{J_{||}S_{j,k}S_{j,k+1}+A(S_{j,k}^{z})^2-g\mu_B B S_{j,k}^{x}\right\}.
\label{hchain}
\end{equation}
The hamiltonian $H_j$ describes the nearest neighbour intrachain interaction
between spins with an easy plane anisotropy ($A>0$) placed in an external magnetic field 
($B$) in the {\bf x} direction. The spins will be treated as classical vectors of lengh $S$ and 
the constants $J_{\|}$ and $J_{\bot}$, both positive, correspond to the antiferromagnetic and 
ferromagnetic exchange coupling constants, respectively. The second term in the r.h.s. of 
(\ref{Htot}) represents an interchain interaction between the spins, completing the description of 
the 3D spin arrangement. Finally, the following values of material parameters will be used: 
$J_{\|}=13.4K$, $S=5/2$, $A/J_{\|}=0.01-0.02$, $J_{\bot}/J_{\|}=1.5 \cdot 10^{-5}$ and $g=2.01$.

In order to start the classical description of the spin dynamics it is convenient to look at two
main different situations, namely, temperatures below and above the transition 
temperature. For temperatures below $T_N$ the system described by (\ref{Htot}) displays
a long range magnetic order, therefore the staggered spontaneous magnetization is not
zero and the system can be described in the mean field approximation as a set of 
non-interacting antiferromagnetic chains with an additional spontaneous magnetization in 
the {\bf y} direction\cite{Holy1}. Explicitly,
\begin{eqnarray}
H&=&\sum_{i}\left\{J_{||}S_{i}S_{i+1}+A(S_{i}^{z})^2-g\mu_B B S_{i}^{x} \right. \nonumber \\
&&\left.-g\mu_B B^{MF}_{\bot}(-1)^{i}S^{y}_{i}\right\},
\label{hmf}
\end{eqnarray}
where
\begin{equation}
B^{MF}_{\bot}=\eta J_{\bot}\langle (-1)^{i}S^{y}_{i} \rangle/g \mu_B,
\label{meanff}
\end{equation}
and $\eta$ accounts for the presence of neighbouring chains in the model. In the specific case of 
TMMC, $\eta=6$. The intrachain mean field $B^{MF}_{\bot}$ is usually replaced by its saturation 
value $B^{S}_{\bot} \approx 22.3$Oe which results from the substitution of $S^{(y)}_{i}$ in
(\ref{meanff}) by its maximum value.

At this point, we can carry on the classical description of the spin dynamics. In order to 
do that it is convenient to change the spin variables to the following form  
\begin{eqnarray}
S_{e,o}&=&\pm S \left[
\sin(\Theta \pm \theta)\cos(\Phi \pm \varphi),\right.\nonumber \\
&&\left.\sin(\Theta \pm \theta)\sin(\Phi \pm \varphi),
\cos(\Theta \pm \theta)
\right],
\label{varia}
\end{eqnarray}
where $e$ and $o$ stands for even and odd sites within a chain.

Using the representation (\ref{varia}) a $\Phi$-dependent part of the hamiltonian (\ref{hmf}) 
can be obtained (see Ref. 3 for details). Explicitly, 
\begin{eqnarray}
H^{\Phi}&=&\frac{1}{2}J_{\|}S^2 \int dz
\left[\frac{1}{c^2}(\partial_t \Phi)^2+(\partial_z \Phi)^2 \right. \nonumber \\
&&\left.-\frac{1}{4}b^2\sin^2 \Phi-2b_{\bot}\sin \Phi 
\right], 
\label{h2sg}
\end{eqnarray}
where
\begin{equation}
c^2=4+\frac{2A}{J_{\|}}, \quad b=\frac{g\mu_{B}B}{J_{\|}S}, \quad 
b_{\bot}=\frac{g\mu_B B^{MF}_{\|}}{J_{\|}S}.
\end{equation}

It should be stressed that the hamiltonian (\ref{h2sg}) is an approximated description of the real
TMMC system. To reproduce the experimental results, magnon-mass and solitonic energy, for instance,
quantum effects and the out of plane component of the magnetization must be taken into 
account.\cite{pires} To go on with the classical description of the $\Phi$-dependent part of the 
original hamiltonian (\ref{Htot}) the equation of motion associated to (\ref{h2sg}), 
\begin{equation}
\frac{1}{c^2}\partial_{tt}\Phi=\partial_{zz}\Phi - \frac{b^2}{8}\sin 2\Phi
-b_{\bot}\sin \Phi 
\label{emot}
\end{equation}
has to be solved. Equation (\ref{emot}) is not completely integrable, however, it has solitonic 
solutions in the form of 2$\pi$-kinks(antikinks) moving with velocity $u$. Explicitly,
\begin{eqnarray}
\cos \Phi &=& \pm2 \frac{\sqrt{\alpha}}{1+\alpha \sinh^2 y} \sinh y,
\label{2sgcs}
\end{eqnarray}
\begin{eqnarray}
\sin \Phi &=& 1-\frac{2}{1+\alpha \sinh^2 y},
\label{2sgss}
\end{eqnarray}
where
\begin{equation}
\alpha=\frac{b_{\bot}}{b_{\bot}+b^2/4}, \qquad
y=(z-z(t))\sqrt{\frac{b_{\bot}+b^2/4}{1-u^2/c^2}},
\label{eye}
\end{equation}
and the position of the soliton center of mass $z(t)$ is given by
\begin{equation}
z(t)=z_0+ut.
\label{eqbali}
\end{equation}

On the other hand, for temperatures above $T_N$ the value of the $b_{\bot}$ is very 
small. In fact, in this situation $b_{\bot}$ can be set equal to zero and, a
well-known solitonic solution for equation (\ref{emot}) can be found: the $\pi$-kink(antikink)
solution for the SG equation
\begin{equation}
\sin\Phi_s (z,t)=\pm\tanh\left[(1-u^2/c^2)^{-1/2}(z-z(t))b/2
\label{sgsol}
\right].
\end{equation}

As it can be seen, the model for TMMC in the continuum approximation leads us to different
kinds of solitonic solutions depending on the temperature. A 2SG soliton solution given by 
(\ref{2sgcs}) and (\ref{2sgss}) for $T<T_N$ and, a SG solution (\ref{sgsol}) for temperatures 
above $T_N$. As it was already mentioned, from the classical point of view, these soliton 
solutions will move with constant velocity througout the sample. However, looking at the
soliton dynamics from the quantum field theory perspective, the interaction with the spin 
waves can transform this ballistic regime into a dissipative one. In the next section
we shall be aiming at the investigation of the 
mobility of the two types of solitons, below and above $T_N$.   

\section{Soliton Mobility}
The quantum dynamics of our spin system (\ref{h2sg}) can be analyzed by studying the
quantum mechanics of the field theory described by the action
\begin{equation}
S[\Phi]=J_{\|}S^{2}\int \int 
\left\{ \frac 1{2c^2}
(\partial_t \Phi )^2-\frac{1}{2} (\partial_z \Phi)^2+U(\Phi )
\right\} dt dz,
\label{action}
\end{equation}
where 
\begin{equation}
U(\Phi)=\frac{b^2}{8}\sin 2\Phi + b_{\bot}\sin \Phi.
\label{2sgpot}
\end{equation} 

To quantize the system described by (\ref{action}) we need to evaluate 
\begin{equation}
G(t)=\text{tr}\int {\cal D}\Phi \ \exp \ \frac i\hbar S[\Phi ]
\label{gt}
\end{equation}
where the functional integral has the same initial and final configurations
and tr means to evaluate it over all such configurations.
As the functional integral in (\ref{gt}) is impossible to be evaluated for a
potential energy density as in (\ref{2sgpot}) we must choose an approximation
to do it. Since the magnetic moments at the manganese sites in the TMMC are large (5/2), 
the semi-classical limit will be chosen as the appropriate one in our case.
Within the functional integral formalism of quantum mechanics, the semi-classical limit
is simply the stationary phase method applied to (\ref{gt}) around the solitonic solutions
(\ref{2sgcs}), (\ref{2sgss}) or (\ref{sgsol}) in which we are interested.
When this is done we are left with an eigenvalue problem that reads
\begin{equation}
\left\{ -\frac{d^2}{dz^2}+U^{\prime \prime }(\Phi _s)\right\} \psi
_n(z-z_0)=k_n^2\psi _n(z-z_0),  
\label{Sle}
\end{equation}
where $\Phi _s$ is denoting the soliton-like solution around which we are
expanding $\Phi (z,t)$ and $\psi _n(z-z_0)$ are the spin wave modes in the presence 
of the soliton.

Now one can easily show that $d\Phi _s/dz$ is a solution of (\ref{Sle}) with $
k_n=0$. The existence of this mode is related to the translation
invariance of the system and causes the divergence of the functional
integral in (\ref{gt}) in the semi-classical limit (Gaussian
approximation).
The way out of this problem is the so-called collective coordinate method 
\cite{christlee}. This method  consists basically in expanding the field
configurations about $\Phi _s(z)$ as

\begin{equation}
\Phi (z,t)=\Phi_s (z-z_0(t))+\sum_{n=1}^\infty c_n\psi _n\left(
z-z_0(t)\right),  
\label{expfi}
\end{equation}
but regarding the $c$-number $z_0$ as a position operator. Using expansion 
(\ref{expfi}), the second quantized version of (\ref{h2sg}) can be written as\cite{Castro}
\begin{eqnarray}
H=\frac 1{2M_s}\ (P-\sum_{mn}\hbar g_{mn}b{_n^{+}}b_m)^2
+\sum \hbar \Omega_nb{_n^{+}}b_n.  
\label{Heff}
\end{eqnarray}
where $\Omega _{n}\equiv ck_{n}$.

In the hamiltonian (\ref{Heff}), $P$ stands for the momentum canonically
conjugated to $z_0$,
\begin{equation}
M_s=\frac {2J_{\|}S^{2}a}{c^{2}}\int_{-\infty }^{+\infty }dzU(\Phi_s(z))  
\label{sm}
\end{equation}
is the soliton mass \cite{rajara} and the coupling constant $g_{mn}$ is
given by
\begin{equation}
g_{mn}=\frac 1{2i}\left[ \sqrt{\frac{\Omega _m}{\Omega _n}}
+\sqrt{\frac{\Omega _n}{\Omega _m}}\right] 
\int dz \psi_m(z) \frac{d\psi _n(z)}{dz}.  
\label{gmn}
\end{equation}

The operators $b^{+}$ and $b$ are respectively the creation and annihilation
operators of the excitations of the magnetic system (magnons) in the
presence of the soliton. In fact, the term
\begin{equation}
\sum_{mn}\hbar g_{mn}b{_n^{+}}b_m,
\end{equation}
can be interpreted as the total linear momentum of the magnons of the system and 
therefore, we are left with a problem in which the momentum associated to the soliton 
is now coupled to the magnons' momenta. 
This effective model suggests that, as the population of magnons is a 
temperature-dependent quantity, the mobility of the soliton will be strongly 
related to the temperature of the system and its dynamics (determined by (\ref{Heff}))
will be non trivial.
 
At this point we are ready to study the mobility
of the wall because we have been able to map that problem into the
hamiltonian (\ref{Heff}), which on its turn has been recently used to study
the mobility of polarons\cite{CN}, heavy particles in 1D environments\cite{CyN} and
skyrmions in 2D electronic systems\cite{villa}. 
The main result obtained in those calculations can be summarized as follows.
The damping function $\gamma (t)$ (basically the inverse of the mobility) is given by 
\begin{eqnarray}
\gamma (t)&=&\frac \hbar {2M}\int_0^\infty \int_0^\infty d\omega d\omega
^{\prime } \left\{ 
S(\omega ,\omega ^{\prime })(\omega -\omega ^{\prime }) \right. \times \nonumber \\
&&\left.[n(\omega)-n(\omega ^{\prime })]\cos (\omega -\omega ^{\prime })t \right\},  
\label{g}
\end{eqnarray}
where
\begin{equation}
n(\omega )=\frac 1{e^{\beta \hbar \omega }-1}
\end{equation}
is the Bose function and,
\begin{equation}
S(\omega ,\omega ^{\prime })=\sum_{mn}|g_{mn}|^2\delta (\omega -\Omega
_n)\delta (\omega ^{\prime }-\Omega _m)  
\label{sww}
\end{equation}
is the so-called scattering function.

In the long time limit $\gamma (t)$ can, to a good approximation, be written as
\begin{equation}
\gamma (t)\cong \bar{\gamma}(T)\delta (t),
\label{gtT}
\end{equation}
where $\delta(t)$ is the Dirac delta function 
and $\bar{\gamma}(T)$ is given by 
\begin{equation}
\bar{\gamma}(T)=\frac 1{2\pi M_s}\int_0^\infty dE\ {\cal R}(E)\frac{\beta E\
e^{\beta E}}{(e^{\beta E}-1)^2}.  
\label{damp}
\end{equation}

In (\ref{damp}), ${\cal R}(E)$ is the reflection coefficient of the ``potential'' 
$U^{\prime \prime }(\Phi _s)$ in the Schr\"{o}dinger-like equation (\ref{Sle}). 
For simplicity we will express the reflection coefficient ${\cal R}(E)$ in 
terms of the even and odd scattering phase shifts\cite{Lipkin} as
\begin{equation}
{\cal R}(k)=\sin ^2\left( \delta ^e(k)-\delta ^o(k)\right).  
\label{Rsin}
\end{equation}
At this point we can perform the calculation of the soliton mobility in TMMC for 
temperatures above and below $T_N$.

\subsection{Soliton mobility for $T<T_N$}
To calculate the soliton mobility below $T_N$ 
we need the explicit form of the potential $U^{\prime \prime }(\Phi _s)$  
involved in (\ref{damp}). For the case of the 2SG soliton this potential can be written 
as\cite{Braun}
\begin{eqnarray}
U^{\prime \prime}(\Phi^{2SG}_s)&=&\frac{1}{\lambda^2}\left[
1-2\mbox{sech}^2(\frac{z}{\lambda}+\rho)
-2\mbox{sech}^2(\frac{z}{\lambda}-\rho) \right. \nonumber \\
&&\left.+2\mbox{sech}(\frac{z}{\lambda}+\rho)
\mbox{sech}(\frac{z}{\lambda}-\rho)\right],
\label{PotV}
\end{eqnarray}
where
\begin{equation}
\lambda=\frac{1}{b_{\bot}+b^2/4}, \qquad \cosh \rho=\frac{1}{\sqrt{\alpha}}.
\end{equation}
The second and third terms in the r.h.s. of (\ref{PotV}) are the potentials
of the noninteracting $\pi$-solitons located at $z/\lambda=\pm \rho$
whereas the last term describes the interaction of the two $\pi$-solitons  
at $z/\lambda=\pm \rho$ respectively. 

For all finite values of $\lambda $ and $\rho$,  
the system is translationally invariant and, consequently, the potential (\ref{PotV})
has a zero energy state that is given by
\begin{equation}
\psi _0\propto
\mbox{sech}(\frac{z}{\lambda}+\rho )+
\mbox{sech}(\frac{z}{\lambda}-\rho ),
\label{fi0g}
\end{equation}
which is nothing but the Goldstone mode of the $2\pi$-soliton for finite transverse 
magnetization and finite external field.

In order to evaluate the expression (\ref{damp}) for the damping constant we
need the even and odd phase shifts associated to the potential (\ref{PotV}).
Unfortunately, their analytical evaluation is very complicated for all finite
values of $\lambda $ and $\rho $, and in what follows we will only study the situation
of weak external fields ($b_{\bot}\gg b^2 /2$).
In this case ($\rho \ll 1$) the Shr\"{o}dinger-like equation
(\ref{Sle}) can be written as
\begin{equation}
\left\{
-\frac{d^2}{dz^2}+V(z) \right\} \psi_{n}(z)=\kappa^{2}_{n}\psi_{n}(z),
\label{fdpu} 
\end{equation}
where
\begin{equation}
\kappa^{2}_{n}=k_n^2-\frac{1}{\lambda^2}-\frac{\rho^{2}}{\lambda^{2}},
\end{equation}
and the potential (\ref{PotV}) is now reduced to the sum of two
contributions, one coming from the spontaneous staggered magnetization and, 
the other from the presence of the weak external field. Explicitly, 
\begin{equation}
V(z)=V_0(z)+(\frac{\rho}{\lambda})^2 V_1(z),  
\label{Vap}
\end{equation}
with
\begin{equation}
V_0(z)=-2\mbox{sech}^2 \left( \frac{z}{\lambda} \right) 
\label{Vap0}
\end{equation}
and
\begin{equation}
V_1(z)=-8\tanh ^2\left(\frac{z}{\lambda}\right)
\mbox{sech}^2 \left( \frac{z}{\lambda} \right) .  
\label{Vap1}
\end{equation}

The calculation of the even and odd phase shifts for a potential of the form 
(\ref{Vap})-(\ref{Vap1}) is reported in Ref. 10 and here we will only show the fundamental results
of the numerical solution of the Schr\"{o}dinger-like equation (\ref{fdpu}).
Fig. 1 and Fig. 2 show the even and odd parity phase shifts for different values
of the external field.
\vskip .20in 
\begin{figure}[h]
\centerline{\psfig{figure=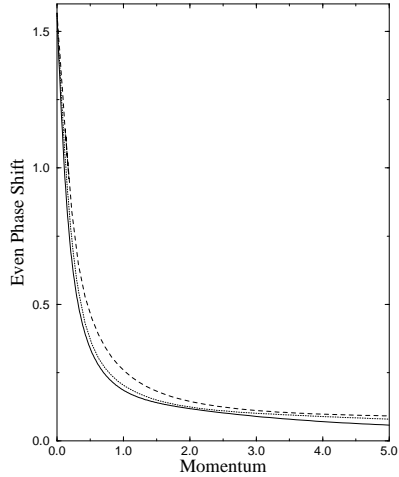,height=2.2in,angle=0}}
\vskip .25in
\caption{The even phase shift as a function of the momentum for different values of $\rho$. 
The continuous line correspond to $\rho=0.14$, the dotted line to $\rho=0.31$ and the dashed 
line to $\rho=0.50$.}
\label{rsb0}
\end{figure}

The values of $\delta_e$ and $\delta_o$ for $k=0$ are in agreement with the 1D version of the 
Levinson's\cite{Bart} theorem which establishes that 
\begin{eqnarray}
\delta ^e(k=0)=\pi (n^e-\frac 12), \nonumber \\
\delta ^o(k=0)=\pi n^o,
\label{levi}
\end{eqnarray}
where $n^e$ and $n^o$ are the number of even and odd parity bound states.
\vskip 0.20in
\begin{figure}[h]
\centerline{\psfig{figure=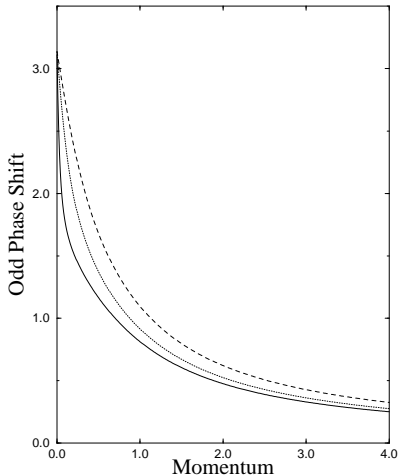,height=2.2in,angle=0}}
\vskip .25in
\caption{The odd phase shift as a function of the momentum. The continuous line correspond to 
$\rho=0.14$, the dotted line to $\rho=0.31$ and the dshed line to $\rho=0.50$}
\label{gapvar}
\end{figure}
As it can be seen in Fig. 1 the even phase shift is $\pi/2$ at the origin. This behavior is in 
complete agreement with the existence of an even bound state corresponding to the Goldstone 
mode. On the other hand, the odd phase shift $\delta_o(0)=\pi$, indicates the presence of 
an odd bound state. This result was previously obtained by Kivshar {\it et al}.\cite{kiv} in the study 
of the small-amplitude modes around the localized solution of the 2SG equation and shows that
there is always an odd bound state in this kind of system.  
Therefore, the spectrum of (\ref{Vap}) is composed by: 
i) the $\psi _0$ solution (\ref{fi0g}) corresponding to the
translation mode of the soliton (Goldstone mode) ii) an internal mode which appears when the
system is perturbed by the external magnetic field and iii) the $\psi _k$ solutions
which constitute the continuum modes and correspond to magnons.

In order to find the damping coefficient we must compute the reflection coefficient ${\cal R}(k)$. 
This can be done by inserting the numerical results of the even and odd phase shifts into the general 
expression (\ref{Rsin}). In Fig. 3, we have plotted ${\cal R}(k)$ for different values of the 
perturbation parameter $\rho$ for the whole range of $k$. As it can be seen
the major contribution for the reflection coefficient comes from the low energy states,
in agreement with the well behaved potentials (\ref{Vap0}) and (\ref{Vap1}).
\vskip 0.2in
\begin{figure}[t]
\centerline{\psfig{figure=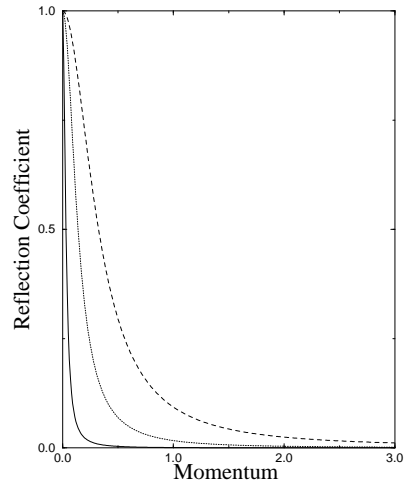,height=2.2in,angle=0}}
\vspace{.25in}
\caption{Reflection coefficient as a function of the momentum. The continuous line corresponds 
to $\rho=0.14$, the dotted line to $\rho=0.31$ and the dashed line is for $\rho=0.50$.}
\label{densityrs}
\end{figure}

Having done that, one can immediately integrate the function ${\cal R}(k)$ in expression 
(\ref{damp})  which finally allows us to describe the damping coefficient as a function of the 
temperature (see Fig.4). It is important to notice that we have not considered the odd bound 
state of the potential (\ref{Vap}) in computing the damping coefficient because in evaluating 
the scattering matrix (\ref{sww}), only elastic terms are taken into account (see for 
instance Ref. 9, Ref. 12 or Ref. 13).
 
As it can be seen, the damping coefficient is linear for high temperatures. This result can be 
obtained directly from (\ref{damp}). In fact, for $T$ high enough the
damping constant can be approximated by
\begin{equation}
\bar{\gamma}(T)\simeq \frac 1{2\pi M_s\beta }\int_0^\infty dE\ \frac{{\cal R}%
(E)}E\propto T  
\label{g1}
\end{equation}
which is linear in $T$, independently of the explicit form of ${\cal R}(E)$. In the 
low temperature regime we can write
\begin{equation}
\bar{\gamma}(T)\simeq \frac 1{2\pi M_s}\int_0^\infty dE\ {\cal R}(E)\beta
Ee^{-\beta E}, 
\label{aoc}
\end{equation}
where $E$ always presents a gap determined by the presence of the magnetic field and/or the 
spontaneous staggered magnetization. Here we shall not attempt to write an approximate expression for 
(\ref{aoc}) because the correct behavior of the reflection coefficient was only numerically 
determined. As it is shown in Fig. 4, for low enough temperatures, the damping coefficient 
drops exponentially to zero due to the existence of the gap.
As the temperature increases the damping coefficient rises following a power law behavior until 
it becomes linear for high enough temperatures. This strong temperature dependence of the damping
parameter, for $T$ below the transition temperature, will influence directly the correlation 
function between the magnetic solitons.

\subsection{Soliton mobility for $T>T_N$}
To perform the calculation of the $\pi$-soliton mobility for $T>T_N$ we simply set to zero the 
$b_{\bot}$ in the hamiltonian (\ref{h2sg}) and therefore, the equation of motion for the 
$\Phi$-dependent part of the spin degree of freedom (\ref{emot}) becomes a SG equation with 
solitonic solution (\ref{sgsol}). In this case the potential involved in the Schr\"{o}dinger-like 
equation (\ref{Sle}) which determines the fluctuations around the soliton solution have the form    
\vskip 0.2in
\begin{figure}[t]
\centerline{\psfig{figure=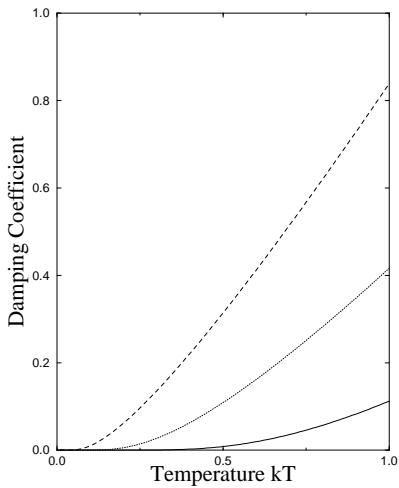,height=2.2in,angle=0}}
\vskip .25in
\caption{The damping coefficient as a function of the temperature for different values of 
$\rho$. The continuous line for $\rho=0.14$, the dashed line for $\rho=0.31$ and the dotted line 
for $\rho=0.50$}
\label{rsb1}
\end{figure}
\begin{equation}
U^{\prime \prime }(z)=\xi ^2(1-2 \mbox{sech}^2 \xi z),
\label{u0}
\end{equation}
where $\xi = b/2$. The spectrum of (\ref{u0}) contains a bound state with zero energy
\begin{equation}
\psi _0=\sqrt{\frac \eta 2}\mbox{sech}
(\xi z),\qquad
k_0^2=0,
\label{gm}
\end{equation}
which constitutes the translation mode of the soliton (Goldstone mode),
and a continuum of quasiparticles modes (magnons) given \cite{MF} by
\begin{equation}
\psi_{n}(x)=\frac{1}{\sqrt{L}}\left[
\frac{k_{n}+i\xi \tanh(\xi z)}{k_{n}+i\xi}
\right]
e^{ik_{n}z},
\label{qp}
\end{equation}
where
\begin{equation}
k_{n}=\frac{2n\pi}{L}-\frac{\delta(k_{n})}{L}, \qquad
\delta(k)=\arctan \left[ \frac{2\xi k}{k^{2}-\xi^{2}} \right].
\label{phase}
\end{equation}

As it was already mentioned, the reflection coefficient ${\cal R}$ for a general
symmetric potential can be expressed in terms of the corresponding even and odd phase
shifts by the relation (\ref{Rsin}). Re-expressing (\ref{qp}) in terms of parity 
eigenstates it is easy to prove that the potential (\ref{u0})  belongs
to the class of reflectionless potentials because its phase shifts are given by
\begin{equation}
\delta ^{e,o}(k)=\arctan (\xi /k),
\label{ps0}
\end{equation}
that do not distinguish between odd and even parities. Therefore no matter how high the temperature
rises above $T_N$ the damping coefficient is always zero and as a direct consequence the
ballistic regime for the soliton results.

As we have seen, the solitonic solutions in TMMC have different regimes for $T$ below and
above $T_N$. Below the transition temperature the $2\pi$-solitons behave like a Brownian particle
with a finite damping parameter. On the other hand, for $T$ above $T_N$ the $\pi$-solitons have 
infinite mobility corresponding to the ballistic regime. The next section is devoted to studying 
the influence of the changes of the solitonic solutions mobility in the dynamical properties of TMMC.  

\section{Dynamical Structure Factor}
In this section we will investigate the dependence of the dynamical properties of TMMC with 
respect to the temperature and the magnetic field. For $T$ below the transition temperature, 
this will be done  through the computation of the dynamical structure factor of a dilute gas of 
2$\pi$-solitons in a dissipative regime. With this result, we can analyze the main differences
with the assumption of ballistic regime using by Holyst\cite{Holy1} in the same situation.

In a general form, the longitudinal and transverse dynamical structure factors with respect 
to the external field $B$ can be defined as 
\begin{equation}
{\mathcal S}^{||(\bot)}=\frac{1}{(2\pi)^2}\int\int dt dz
e^{i(qz-\omega t)}\langle S^{x(y)}(0,0)S^{x(y)}(z,t)\rangle,
\end{equation}
where $S^{x(y)}$ corresponds to the spin component in the {\bf x}({\bf y}) direction.
  
To begin with, let us recall the main results for the longitudinal dynamical structure factor
reported in Ref. 3. Using the model of non-interacting 2$\pi$-soliton gas in the ballistic regime
${\mathcal S}^{||}(q,\omega)$ can be written approximately as
\begin{equation}
{\mathcal S}^{||}(q,\omega)=n_{2\pi}S^2|F^{x}_{2\pi}(q)|^2 \frac{p(\omega/q)}{2\pi q},
\label{corr}
\end{equation}
where
\begin{equation}
p(\omega/q)=\sqrt{\frac{\beta E_{2\pi}}{2\pi c^2}}
\exp{-\frac{\beta E_{2\pi}\omega^2}{2c^2 q^2}},
\label{gau1}
\end{equation}
\begin{eqnarray}
E_{2\pi}&=&2Bg\mu_B S \left[
\sqrt{1+4b_{\bot}/b^2} \right. \nonumber \\
&&\left. +4b_{\bot}b^{-2}\sinh^{-1}(bb^{-1/2}_{\bot}/2)\right]
\end{eqnarray}
and
\begin{equation}
F^{x}_{2\pi}(q)=\frac{i \pi d_{\pi}}{2}\frac{\sin (qd_{\pi}\sigma)}
{\cosh (q\pi d_{\pi}\sqrt{1-\alpha}/8)}, 
\end{equation}
\begin{equation}
\sigma=\frac{\sqrt{1-\alpha}}{8}\ln\left(\frac{2}{\alpha}-1+
\frac{2}{\alpha}\sqrt{1-\alpha}\right), \qquad d_{\pi}=\frac{8}{b}.
\label{defi}
\end{equation}
The correlations described by (\ref{corr}) are induced by single kinks moving from the
origin to the position $z$ in a time interval 0 to $t$. As expected, the Maxellian velocity
distribution used to describe the $2\pi$-kink gas is directly reflected in the Gaussian 
dependence of the longitudinal structure factor with the frequency. 

On the other hand, to get a better idea of the changes in the dynamical properties when we 
cross the transition temperature, it is convenient to calculate the dynamical structure factor 
for $T$ above the transition temperature. As it was demostrated before, above $T_N$, the 
$\pi$-solitons moves without dissipation and, therefore, the ballistic regime is valid. Using again 
the model of a dilute gas of solitons, the dynamical structure factor can be written as
\begin{eqnarray}
{\mathcal S}^{||}(q,\omega)&=&\frac{S^2}{(2\pi)^{3/2}}
\sqrt{\frac{E_{\pi}\beta}{c^2 q^2}}|F^{\|}(q)|^2\exp(-\frac{E_{\pi}\beta\omega^2}{2 c^2 q^2}),
\label{gau2}
\end{eqnarray}
where
\begin{equation}
F^{\|}(q)=\frac{2\pi}{b}\mbox{sech}(q\pi/b) \qquad \mbox{and} \qquad E_{\pi}=Bg\mu_BS.
\end{equation}
As it can be seen, the dependence with the frequency remains almost unchanged no matter what the 
temperature is. At the same time, it should be noticed that once the intensity of the central peak 
and the density of kinks in (\ref{gau1}) and (\ref{gau2}) are proportional, the intensity of the 
central peak for $T<T_N$ will be lower. This is a consequence of the smaller number of solitons 
for temperatures below the transition temperature.\cite{Holy1}

With the previous results for $T$ above and below $T_N$ in mind, we can go further on and 
study the influence of the dissipative regime in the dynamical properties of the $2\pi$-kink gas.
As it was shown before, below the transition temperature the $2\pi$-solitons move in a 
dissipative regime. Therefore, the position of the center of mass for those kind of excitations 
as a function of time can be written as
\begin{equation}
z(t)=z_0+\frac{v_0}{\gamma(T)}(1-\exp-\gamma(T) t),
\label{dizzi}
\end{equation}
where $z_0$ is the initial position, $v_0$ is the initial velocity and $\gamma(T)$ is the
temperature-dependent damping parameter. Now, to calculate the dynamical structure factor we
will use the $2\pi$-soliton solutions (\ref{2sgcs})-(\ref{eye}) with $z(t)$ given by 
(\ref{dizzi}). Following the same procedure that led to equation (\ref{corr}) and, after some 
calculations, ${\mathcal S}^{||}(q,\omega)$ can be written as
\begin{equation}
{\mathcal S}^{||}(q,\omega)=\frac{2n_{2\pi}S^2}{\pi}|F^{x}_{2\pi}(q)|^2 \Gamma(q,\omega),
\label{corr2}
\end{equation}
where
\begin{equation}
\Gamma=\sum_{n=0}^{\infty} \frac{(-1)^n}{n!}
\left( \frac{q^2}{2\beta E_{2\pi} \gamma^2} \right)^n
\sum_{m=0}^{2n}(-1)^m C^{2n}_{m}\frac{2 m \gamma}{m^2\gamma^2+\omega^2}
\end{equation} 
and
\begin{equation}
C^{2n}_{m}=\frac{(2n)!}{m!(2n-m)!}.
\label{fact}
\end{equation}

As it can be seen, the dissipative regime for the magnetic solitons in the case in which $T<T_N$, 
changes considerably the behavior of the longitudinal dynamical structure factor. Although the 
expression (\ref{corr2}) is valid for all finite values of $q$, we couldn't perform the entire sum
to get a closed expression. Therefore, it is helpful to study the behavior of 
(\ref{corr2})-(\ref{fact}) for small momentum in order to compare it with the ballistic behavior
result (\ref{corr}). Assuming that
\begin{equation}
q\ll \frac{\gamma}{c}\sqrt{2\beta E_{2\pi}},
\end{equation} 
the dynamical structure factor ${\mathcal S}^{||}(q,\omega)$ can be written as
\begin{equation}
{\mathcal S}^{||}(q,\omega)=\frac{2\alpha}{\pi}|F^{x}_{2\pi}(q)|^2 
\Lambda(q,\omega)
\end{equation}
where
\begin{eqnarray}
\Lambda(q,\omega)&=&2\pi \delta(\omega) \exp-\frac{q^2 c^2}{\beta E_{2\pi} \gamma^2} + \nonumber \\
&&\frac{q^2 c^2}{\beta E_{2\pi} \gamma^2}\left[
\frac{\gamma}{\gamma^2+\omega^2}-\frac{\gamma}{4\gamma^2+\omega^2}
\right].
\label{loren}
\end{eqnarray}

Within the approximation of small momentum, the behavior of  
${\mathcal S}^{||}(q,\omega)$ with the frequency, changes from the `Gaussian' central peak to a 
`Lorentzian' dependence. Therefore, as the temperature is lowered below $T_N$, the central peak 
behavior is replaced by a smoother-one in the frequency domain. This result is a direct consequence 
of the dissipative regime of the $2\pi$-soliton and, as the damping constant $\gamma$ can be controlled 
by changing  the temperature and the magnetic field, a possible indication of a non-ballistic regime 
has been found. Another quantity that can be computed in order to get a better idea of the influence 
of the dissipative motion of magnetic solitons in TMMC is the $T_1$ time of NMR. This problem is 
currently being investigated by one of us.

\section{Conclusions}
In this paper we have analyzed the possibility of identifying two different regimes of motion for 
the magnetic solitons in the TMMC antiferromagnet. We were able to show that above the transition 
temperature $T_N$, the $\pi$-soliton moves without dissipation, even
from the field theoretical point of view. This result is in complete agreement with the 
ballistic regime adopted to understand the experimental data reported in Ref. 11. 

On the other hand, for $T$ below the transition temperature the $2\pi$-solitons 
in the system move with finite mobility. Therefore, a dissipative equation of motion
has to be used in the description of the soliton's center of mass motion. This difference in the 
regime of motion is directly reflected in the longitudinal structure factor and, therefore,
can be used as an indication of the finite mobility of the solitonic solutions below the transition 
temperature. The results presented here could be directly compared to the experimental data
one may obtain when testing the TMMC antiferromagnet in this temperature regime.

Although the formulation used to compute the damping parameter is valid for all values of the
external magnetic field, we have restricted ourselves to the study of very weak fields. However, 
our formulation can be used to study situations with arbitrarily stronger magnetic
fields and to other magnetic materials that support solitonic solutions without any major qualitative
difference. For instance, we could treat systems modelled by the 1-D Dzyaloshinski-Moriya 
antiferromagnet which can naturally be described by a 2DSG\cite{dzmo} hamiltonian independently of the 
temperature.

\section{Acknowledgment}
AVF wishes to thank Funda\c{c}\~{a}o de Amparo \`a Pesquisa do Estado de S\~{a}o Paulo 
(FAPESP) for financial support, whereas AOC kindly acknowledges partial
support from Conselho Nacional de Desenvolvimento Cient\'{\i}fico e Tecnol\'ogico (CNPq).

\end{multicols}

\end{document}